\documentclass[runningheads]{cl2emult}

\usepackage{makeidx}  
\usepackage{graphicx} 
\usepackage{subeqnar} 
\usepackage{multicol} 
\usepackage{cropmark} 
\usepackage{eso}      
\makeindex            

\newcommand{\Msun}{\mbox{$M_{\odot}$}}
\newcommand{\ga}{\stackrel{>}{_{\sim}}}
\newcommand{\la}{\stackrel{<}{_{\sim}}} 

\begin{document}
\title*{Fine structure of the red clump\protect\newline in Local
Group galaxies}
\toctitle{Fine structure of the red clump\protect\newline in Local
Group galaxies}
%
%
\titlerunning{The red clump in galaxies}
%
\author{L\'eo Girardi\inst{1,2}}
\authorrunning{L.\ Girardi}
%
%
\institute{Dipartimento di Astronomia, Vicolo dell'Osservatorio~5, 
	I-35122 Padova, Italy
\and Max-Planck-Institut f\"ur Astrophysik, Karl-Schwarzschild-Str.~1,
	D-85740 Garching bei M\"unchen, Deutschland}

\maketitle              

\begin{abstract}

\index{abstract}Some fine structures can nowadays be identified in 
the high-quality colour-magnitude diagrams (CMD) of Local Group galaxies. 
The clump of red giants, for instance, 
may present a significant colour spread,
and extensions to both brighter 
and fainter luminosities. Such features are predicted by population 
synthesis models which consider stars in the complete relevant ranges of 
ages and metallicities, and are potentially useful for constraining 
the star formation histories of the parent galaxies over scales of 
gigayears. We briefly 
comment the cases of fields in the Magellanic Clouds, M31, 
and the local CMD from {\em Hipparcos}. 
\end{abstract}

\subsection*{Introduction}
The red clump is a striking feature in the CMDs of intermediate-age and
old open clusters, and in those of the nearest galaxies. It is composed 
mostly by low-mass stars in the stage of core helium burning (CHeB). 
In these stars, the onset of electron 
degeneracy after the central H-exhaustion postpones the He-ignition until
the core mass grows to about $M_{\rm c}\simeq0.45$~\Msun. It causes all
low-mass CHeB stars to have similar luminosities and hence position 
in the CMD.

In open clusters, the red clump typically presents a very small dispersion
in both colour and magnitude. For instance, in M~67 the red clump is defined 
by 6 stars with $\sigma(B-V)\la0.05$, $\sigma(V)\la0.1$~mag (see e.g.\
Montogomery et al.\ 1993). This 
results from the small spread of masses and metallicities among these stars. 
On the contrary, red clump stars in galaxy fields  
present a significant spread in both mass and metallicity; 
as a consequence, these clumps are expected to present a fine
structure in the CMD (Girardi 1999).
\begin{figure}
\resizebox{0.55\textwidth}{!}{\includegraphics{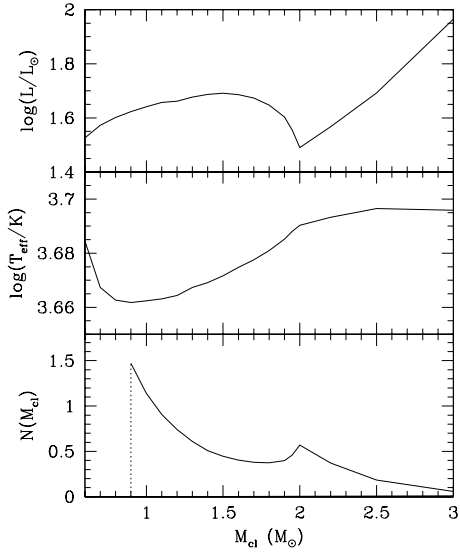}}
\hfill
\parbox[b]{0.4\textwidth}{
\caption[]{For $Z=0.019$, we present the mean luminosity (upper panel)
and effective temperature (middle) of CHeB stars as a function of mass.
The lower panel shows their mass distribution for a galaxy 
model with constant star formation rate up to 10 Gyr ago.}
\label{fig_massdist}
}
\end{figure}

This fine structure is not evident in the case of most
ground-based observations of nearby galaxies, 
for which the clump spread in the CMD reflects, to a large
extent, the presence of large photometric errors. 
The {\em Hipparcos} satellite, however, has provided 
an impressive CMD of the nearby stars (Perryman et al.\ 1997), 
in which the red clump clearly presents an intrinsic structure 
(Girardi et al.\ 1998). HST has also provided
very nice CMDs for the Magellanic Clouds (MC) and other nearby 
dwarf galaxies, in which sub-structures of the clump seem to be present.
In the case of the MCs, however, too few
clump stars can be sampled in a typical WFPC2 frame,
so that the clump features are not so clear. 
This situation is expected to improve with the FORS camera at VLT 
since it can provide CMDs with quality comparable to the HST ones
(at least for the Magellanic Clouds), but sampling 
much larger fields.

\subsection*{Theoretical expectations}
In order to simulate the clump structure in different galaxy 
models, we make use of the extensive set of evolutionary tracks 
and isochrones of
Girardi et al.\ (1999). This database covers a large interval
of stellar initial masses (from 0.15 to 7~\Msun) and metallicities 
(from $Z=0.001$ to 0.03), with a very good mass resolution. 
Synthetic CMDs are then generated by means
of a population synthesis tool, for any arbitrary history of 
star formation and chemical enrichment.

Fig.~\ref{fig_massdist} 
illustrates the location of solar metallicity ($Z=0.019$) CHeB
stars in the HR diagram, as a function of mass, 
and their mass distribution in a 
model galaxy which formed stars at a constant rate from 0.1 to 10 Gyr ago,
with a Salpeter IMF and with Reimers' mass-loss rates along the RGB
(see Girardi 1999 for details).
The following aspects are evident in this figure:
	\begin{itemize}
\item stars with 
$M\la2$~\Msun\ (i.e.\ low-mass stars) constitute most of the
clump, distribute over a non-negligible range in $T_{\rm eff}$, 
and with an almost 
constant luminosity. They form the {\em main red clump} feature
we are used to.
\item stars with $M\simeq2$~\Msun\ 
not only ocupy a particular region of the HR diagram (they are about 
0.4~mag fainter, and slightly bluer than most clump stars), but 
represent a second peak in the mass distribution. Therefore, these stars
define a fainter {\em secondary red clump}.
	\begin{figure}
\resizebox{0.55\textwidth}{!}{\includegraphics{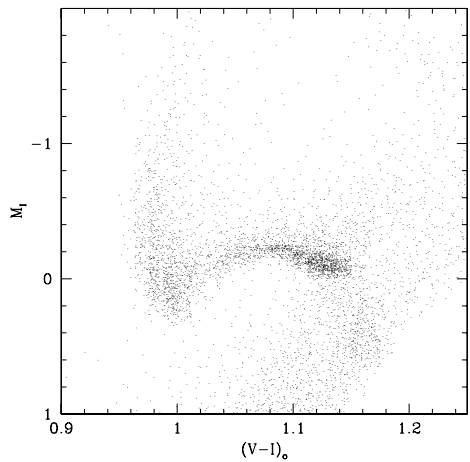}}
\hfill
\parbox[b]{0.4\textwidth}{
\caption[]{Synthetic $M_I$ vs.\ $V-I$ 
CMD for a galaxy model which assumes $Z=0.019$
(with a small metallicity dispersion), and constant star formation rate. 
Only a small region of the CMD, centered on the red clump, is shown.}
\label{fig_cmd}
}
	\end{figure}
\item stars with $M\ga2$~\Msun\ (i.e.\ intermediate-mass 
stars) appear at higher luminosities
and in non-negligible quantities. They may originate a plume of bright 
clump stars (or the so-called {\em vertical red clump}).
	\end{itemize}
These structures are evident in the synthetic CMD of 
Fig.~\ref{fig_cmd}. We remark that essentially the same features 
appear at different metallicities, provided that $Z\ga0.004$. For lower
metallicities, the colour spread of the clump gets very low, unless 
for the tail of lowest-mass clump stars, which may distribute
over a blue horizontal branch if ages are high enough.
On the other hand, models which assume an intrinsic age-metallicity
relation for clump stars, tend to present the main clump distributed
over a large colour interval, but still at an almost constant luminosity
(Girardi 1999). 

\subsection*{Observations and pratical implications}
From the above-mentioned work, it results that the clump structure
in the CMD contains potentially useful information about the 
history of star formation and chemical enrichment of the parent galaxy.
The most evident example is given by
the secondary clump stars in a 
field: their number should be simply proportional to the SFR at an age of 
$\sim1$~Gyr 
ago (corresponding to an initial mass of 2~\Msun\ in the present models). 
The colour distribution of the main clump, instead, should reflect 
both the age and metallicity distribution of clump stars, 
over timescales of gigayears.
Importantly, this information appears at a luminosity level 
which is at least 3~mag brighter than the oldest turn-offs of most
galaxies.

Also, it is important to emphasise that the model predictions are 
supported by a number of interesting observations. For instance, a 
secondary clump has been noticed in the {\em Hipparcos} CMD of local 
stars (Girardi et al.\ 1998) and in some LMC fields observed by Bica et 
al.\ (1998). Also, Corsi et al.\ (1994) observed 
a minimum in the clump luminosities for LMC clusters of ages $\sim1$~Gyr,
which corresponds to the secondary clump we mention here.

Other important observation which can be accounted for 
by present models is the nearly constancy of the clump $I$-band
luminosity, $M_I$, with the $V-I$ colour, 
noticed by Paczy\'nski \& Stanek (1998), 
Stanek \& Garnavich (1998),
Udalski et al.\ (1998) and Stanek et al.\ (1998). This was 
interpreted as indication that the clump luminosity does not
depend on the stellar population. If this were the case, the red clump
would be an excelent standard candle, 
with a zero point precisely determined by means of
{\em Hipparcos} data.
Theoretical models instead indicate a non-negligible dependence of
$M_I$ on age and metallicity, amounting to a maximum of 0.6~mag
(see also Cole 1998). At first sight, it
seems in contradition with the observations. 
However, constant $M_I$ as a function of $V-I$ can easily be obtained 
in population synthesis models, if we assume either a 
nearly constant $Z$ with varying age, or a $Z$ decreasing
with age according to a normal age--metallicity relation 
(see Girardi et al.\ 
1998 and Girardi 1999, for details). The first case seems to describe 
reasonably what observed in the {\em Hipparcos} CMD, 
whereas the second case
would correspond better to what observed in M~31. 

It follows that the red clump 
may be a useful distance indicator only if we know the distribution
of ages and metallicities inside the galaxy we are observing. 
Otherwise, systematic errors as large as $\sim0.4$~mag may be present 
in the distance modulus derived by means of this method.

\paragraph{Acknowledgements}
Part of this work has been carried out at the Max-Planck-Institut f\"ur
Astrophysik during a stay funded by the Alexander
von Humboldt-Stiftung. The collaboration with A.\ Weiss, M.A.T.\ 
Groenewegen and M.\ Salaris was decisive during this period. I thank 
ESO for a travel grant to attend the VLT opening symposium.

\end{document}